\documentclass[aps,prl,reprint,floatfix,nobibnotes,superscriptaddress]{revtex4-2}

\usepackage{color} 
\usepackage{bm} 
\usepackage{graphicx} 
\usepackage[T1]{fontenc}
\begin{document}
\title{Steering Active-Colloid Assembly by Biasing Dissipation}

\author{Chaoqun Du}
\affiliation{Department of Chemical Physics, University of Science and Technology of China, Hefei, Anhui 230026, China}

\author{Zhiyu Cao}
\email{zc61@rice.edu}
\affiliation{Center for Theoretical Biological Physics,
Rice University, Houston, TX 77005, USA}

\author{Zhonghuai Hou}
\email{hzhlj@ustc.edu.cn}
\affiliation{Department of Chemical Physics, University of Science and Technology of China, Hefei, Anhui 230026, China}
\affiliation{Hefei National Research Center for Physical Sciences at the Microscale}
\date{\today}

\begin{abstract}
Complex nonequilibrium self-assembly enables the formation of materials with specific patterns and functions from the bottom up. How to directionally control the assembly to form the target configuration is a challenge. Here, we propose a dissipation bias principle for targeted assembly, which highlights that controlling the dissipation tendency can play an important role by modulating the frequency and intensity of local rearrangements. Following this principle, one can induce ordered target configurations from disordered structures and also achieve directional selection among multiple assembly pathways. We use the assembly of active colloids as a platform to show our results.
\end{abstract}

\maketitle

\textit{Introduction.}---Self-assembly refers to the autonomous organization of components into patterns or structures, which is an ideal approach to achieving complex organization and an important research subject in biochemistry \cite{whitesides2002self}. In complex self-assembly, precise control over interactions among components is required to achieve the desired structures. Such control includes tuning the concentration of components, as well as the specificity and strength of interactions, among other factors. Additionally, due to the dynamic nature of the system, interactions between components may lead to heterogeneity or competitive phenomena, thereby influencing the final structure and performance. In equilibrium states, even though target structures are encoded through specific interactions \cite{erb2009magnetic,sacanna2010lock,chen2011directed,wang2012colloids,damasceno2012predictive,manoharan2015colloidal}, the system can still easily fall into kinetic trapping states with numerous defects, especially in high-density self-assembly systems \cite{rey2017anisotropic,bodnarchuk2011structural}. Therefore, the dynamic control of targeted self-assembly processes remains a major challenge.

A common strategy widely used in nature is to bypass the entropy bottleneck through energy input, which can assemble features that are impossible to appear in equilibrium states, and it can even be multi-target states \cite{bisker2018nonequilibrium}. However, without control over the dissipative trends, the system still tends to form random components that are not adapted to the environment \cite{boekhoven2015transient,carnall2010mechanosensitive,sadownik2016diversification,tena2017non}. Here, we employed a biased ensemble approach to demonstrate that controlling the dissipation tendency quantitatively constitutes a dissipation bias principle for complex self-assembly.  The biased ensemble approach is built based on the stochastic thermodynamics \cite{seifert2012stochastic,tociu2019dissipation,pietzonka2019autonomous,cagnetta2017large,nemoto2019optimizing,fodor2020dissipation,fodor2022irreversibility,o2022time} and large deviation theory \cite{tociu2019dissipation,cagnetta2017large,nemoto2019optimizing,das2021variational,touchette2009large,fodor2020dissipation,jack2020ergodicity,o2022time,fodor2022irreversibility,lamtyugina2022thermodynamic}, regularly monitoring the dissipation during the assembly process of replicas. This approach can effectively capture energy-avoiding/seeking paths and gradually collect the energy-avoiding/seeking states, ultimately linking to the targeted configuration. At the statistical level, controlling the dissipation tendency regulates the frequency and intensity of local structural rearrangements, which leads to renormalized interactions. Following the principle we proposed,
one can not only induce the emergence of targeted ordered configurations from disordered structures but also directionally select targeted configurations from multiple assembly pathways. We demonstrated our dissipation bias principle using the assembly of active colloids as a platform.

%Here, we investigate the control of self-assembly through dissipation, using the active core-corona particles (ACCPs) as a platform.  In the ACCPs model, without non-equilibrium drivings, the repulsion between colloidal soft cores significantly inhibits the rearrangement of local structures, leading the system into a disordered state.  We propose a thermodynamic control principle to regulate the frequency and intensity of local structural rearrangements in colloids, enabling the generation of trajectories exhibiting different microscopic structural features of energy-avoiding/seeking processes.  This approach, unlike traditional methods that directly adjust the magnitude of active drivings, utilizes thermodynamic control to gradually induce the emergence of ordered configurations from disordered structures and enhance the stability of ordered configurations.  More interestingly, we show that the multi-stable assemblies can be selectively modulated through the quantitative control of thermodynamic inputs, driving colloids to form distinct highly concentrated ordered structures, which provides a general strategy for directed assembly.

\begin{figure}[htbp]
\includegraphics[width=0.98\columnwidth]{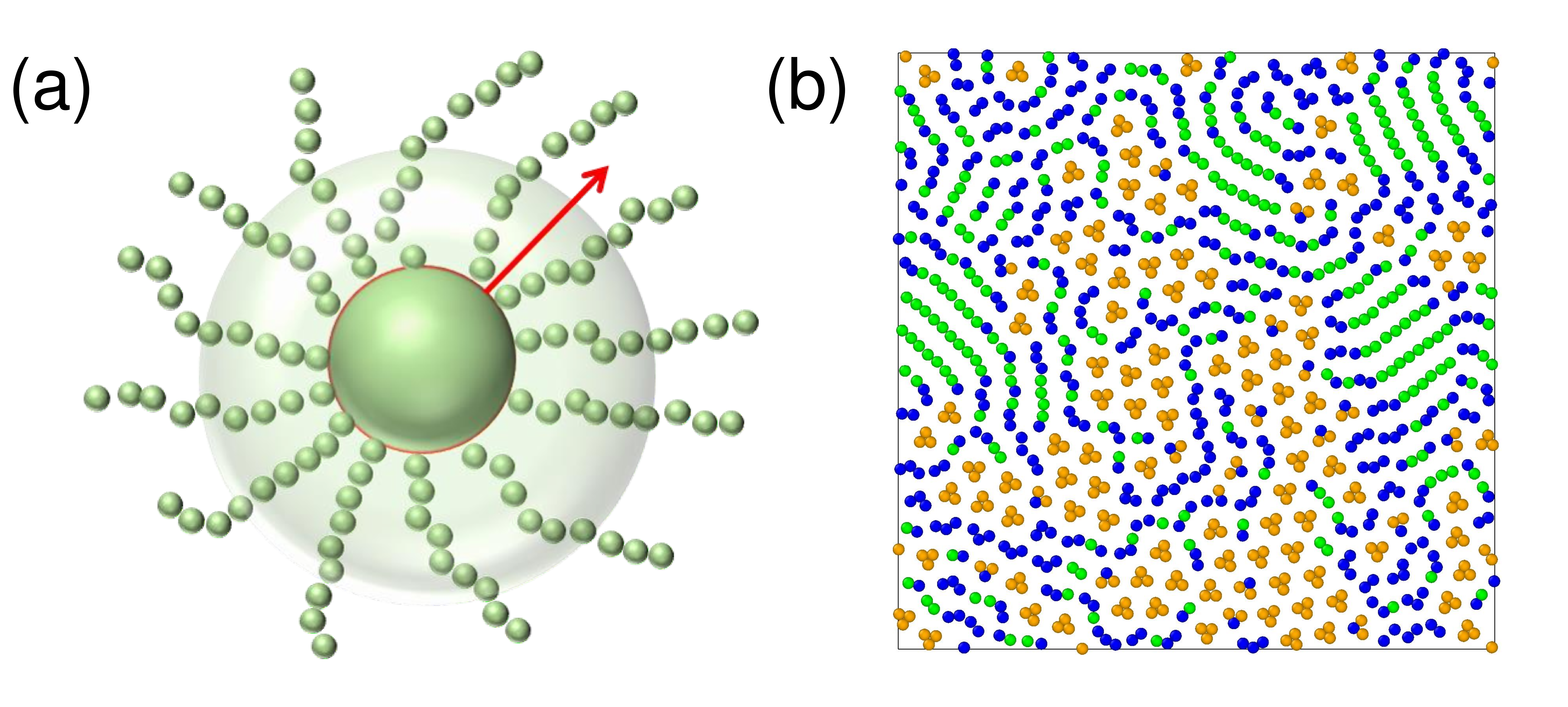}
\caption{(a) The sketch of the model of the active core-corona particles (ACCPs). (b) The typical configuration of the ACCP assembly system with the disordered phase (blue), stripe phase (green), and trimer phase (orange) in the steady states.}
\label{fig:1} 
\end{figure}

\textit{Model.}---First, we reproduced the stripe phase and trimer phase observed in experiments \cite{seul1995domain,stoycheva2000stripe,malescio2003stripe} through the molecular dynamics simulation of $N$ active core-corona particles (ACCPs) in two dimensions (2D). The so-called core-corona colloids are a type of particle with a nanoparticle core and polymer or gel coatings on the surface \cite{si2018nanoparticle}, which are capable of producing rich structures with unusual symmetry \cite{malescio2003stripe,dotera2014mosaic,portehault2008core,jagla1998phase,du2019self}.  Fig.~\ref{fig:1}a shows a schematic diagram of the ACCP particle. The core with a diameter of $\sigma$ simulates the impenetrable hardcore of the ACCP. Its interactions are described by the Weeks-Chandler-Andersen (WCA) interaction potential: $U_{core}(r_{ij})=4\varepsilon[(\sigma/r_{ij})^{12}-(\sigma/r_{ij})^{6}+1/4]$ for $r_{ij}<2^{1/6}\sigma$, and $U_{core}(r_{ij})=0$ otherwise, where $\varepsilon$ is the interaction strength. As for the outer shell, we use the Hookean springs to simulate its interactions \cite{cheng2010probing}, given by $U_{corona}(r_{ij})=k_s(r_{ij}-r_c)^2/2$ for $r_{ij}<r_c$, and $U_{corona}(r_{ij})=0$ otherwise, similar to the soft repulsion between soft-core particles formed by microgels \cite{rey2017anisotropic}. $k_s$ is the strength of the spring potential and $r_c$ is the cutoff distance. Each colloid undergoes Brownian dynamics at a constant temperature $T$ described by the coupled overdamped equations:
\begin{equation}
\dot{\bm {r}}_{i}=\mu \bm{F}_i+v_0 \bm{u}(\theta_i)+\sqrt {2D_t}\bm{\eta_i}; \qquad \dot{\theta}_i =\sqrt {2D_r}\xi_i,\label{eq:LE}
\end{equation}
where $\mu$ represents the particle's mobility, and $\bm{\eta}_i$, $\xi_i$ are zero-mean unit variance Gaussian white noises. $D_t$, $D_r$ are the translational and rotational diffusivities. The active driving term is introduced through a self-propulsion speed of constant magnitude $v_0$ and is modeled along a predefined orientation vector $\bm{u}(\theta_i)=(\cos\theta_i,\sin\theta_i)$, which passes through the particle's center of mass.

The system in the simulation exhibits different structures for different values of the spring potential strengths  ($k_s$) and the self-propulsion speed ($v_0$). Fig.~\ref{fig:1}b shows three characteristic configurations that exist in the parameter space: the disordered phase (blue), the stripe phase (green), and the trimer phase (orange). We differentiate among these characteristic configurations based on the angle $\theta$ ($0^\circ < \theta < 180^\circ$) between a particle and its two adjacent particles. If this angle exceeds $160^\circ$, we classify the particle as belonging to the stripe phase; if the angle falls within the interval ($50^\circ$, $70^\circ$), we classify the particle as belonging to the trimer phase; particles falling outside these two types are classified as disordered. If a particle does not have exactly two adjacent particles or has more than two adjacent particles, it is also classified as disordered.

\textit{Dissipation bias principle.}--- For active colloids, dissipation can be characterized by the averaged power input of the self-propelled force \cite{cagnetta2017large,pietzonka2019autonomous,nemoto2019optimizing}:
\begin{equation}
\dot{\omega}=\frac{1}{N\tau}\sum_i^N\int_{0}^{\tau}\dot{w}_idt=\frac {v_0}{N\mu \tau}\sum_{i=1}^{N}\int^{\tau}_{0}\dot{\bm {r}}_{i}\circ\bm{u}(\theta_i)dt,\label{eq:AW}
\end{equation}
which naturally measures how efficiently active components create motion through energy consumption. Here, $\tau$ is the time duration, and "$\circ$" represents the Stratonovich convention.

\begin{figure}[htbp]
\includegraphics[width=0.98\columnwidth]{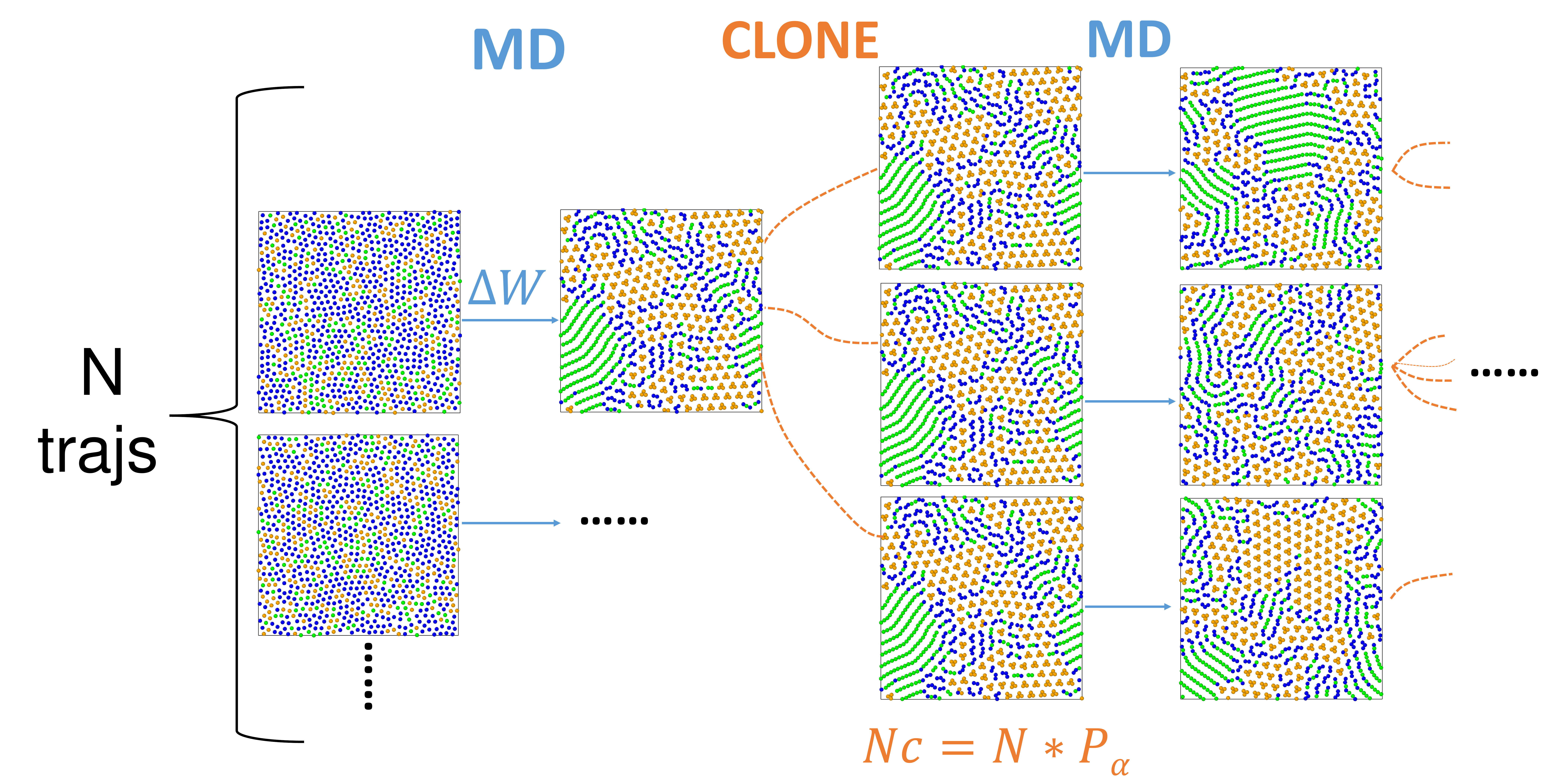}
\caption{Schematic of the cloning algorithm for dissipation-biased trajectory sampling. A population of $N$ trajectories is propagated in parallel by unbiased MD for a time window $\tau$. At the end of each window, each trajectory is assigned a statistical weight $P_{\alpha}$ determined by the dissipation $\Delta W$ accumulated over $\tau$. The population is then resampled by cloning or pruning trajectories in proportion to $P_{\alpha}$ and the number of clones is $N_c$. 
}
\label{fig:2} 
\end{figure}

%In \textcolor{red}{Fig.\ref{fig:1}(b)}, we depicted the phase diagram of the active power input $\dot{\omega}$ with $k_s$ and $v_0$ by using the Eq. \ref{eq:AW}. When no active power is applied, the repulsion of the surface layer seriously inhibits the adjustment of the local structure, and the system quickly relaxes to a steady state with a large number of defects. At this time, the proportion of the three phases essentially remains constant over time, and the disordered structure dominates. In other words, the system is difficult to assemble into ordered structures, whether stripe structures or trimer structures, without the assistance of power input through active driving. First, to reveal how thermodynamic power input assists the assembly, we investigate the effect of different active power inputs on assembly under fixed interactions. Interestingly, we find that there is an optimal intermediate active power input that maximizes assembly state components while minimizing defect structures, as shown in \textcolor{red}{Fig.\ref{fig:1}(c)  for $k_s=200$ and (d) for $k_s=300$}. At this point, the proportions of stripe or trimer structures can exceed $0.9$, while the fraction of another assembly state is nearly zero. This demonstrates that active input can assist core corona particles in self-assembling into highly ordered structures. Thermodynamic control principle to enhance the yield and stability. The free energy profiles of the system over the reaction coordinate $\phi_{tri}$.

 Here, we find that the target state is sensitive to the power input, systems are usually disordered when the dissipation tendency is not controlled. On the one hand, when the active power input is too small, the repulsion of the surface layer strongly inhibits the adjustment of the local structure. The motion of clusters formed by a small number of particles with low energy barriers, which act as quasi-particles, leads to spatially heterogeneous dynamics, and the system quickly relaxes to a steady state with a large number of defects \cite{nishikawa2024collective}. Thus, the disordered structure dominates in this regime. On the other hand, when the active power input is too large, the system is more inclined to form isotropic disordered structures rather than the targeted ordered state because of the excessively intense local rearrangements. Therefore, there is an optimal dissipation for the assembly state to remain stable. At the statistical level, the optimal dissipation for maximizing the yield of targeted states arises from a moderate level of local rearrangements, which requires controlling the system's dissipation tendency(see Fig.~S2).

The large deviation theory and biased sampling approaches \cite{garrahan2007dynamical,hedges2009dynamic,pitard2011dynamic,speck2012large,chetrite2013nonequilibrium} offer us a framework to achieve control over the dissipation tendency in the simulation. By choosing the trajectory-dependent active power input as the observable, we employ the cloning algorithm to generate biased trajectories \cite{tailleur2007probing,nemoto2016population}, as schematically illustrated in Fig.~\ref{fig:2}. The probability of biased trajectories is given by 
\begin{equation}
P_{\alpha}\propto P_0 e^{\alpha \dot{\omega} \tau}.\label{eq:Pa}
\end{equation}
Here, $P_0$ is the probability of unbiased trajectories. We use the parameter $\alpha$ to bias sample trajectories according to active power input, thereby controlling the system's dissipation tendency. Positive values of $\alpha$ sample energy-seeking paths with high dissipation during the assembly process, while negative values sample energy-avoiding paths with low dissipation during the assembly process. By monitoring the system's response to the control of dissipation tendency at fixed periods, we can obtain trajectories corresponding to pumping active power into colloidal units or extracting energy from them at a specific rate. More detailed description of the biased ensemble approach can be found in the supplementary materials (see SM). At the statistical level, controlling the dissipation tendency modulates the intensity and frequency of the local structure rearrangement, which effectively renormalizes the interaction. 

\begin{figure}[htbp]
\includegraphics[width=0.98\columnwidth]{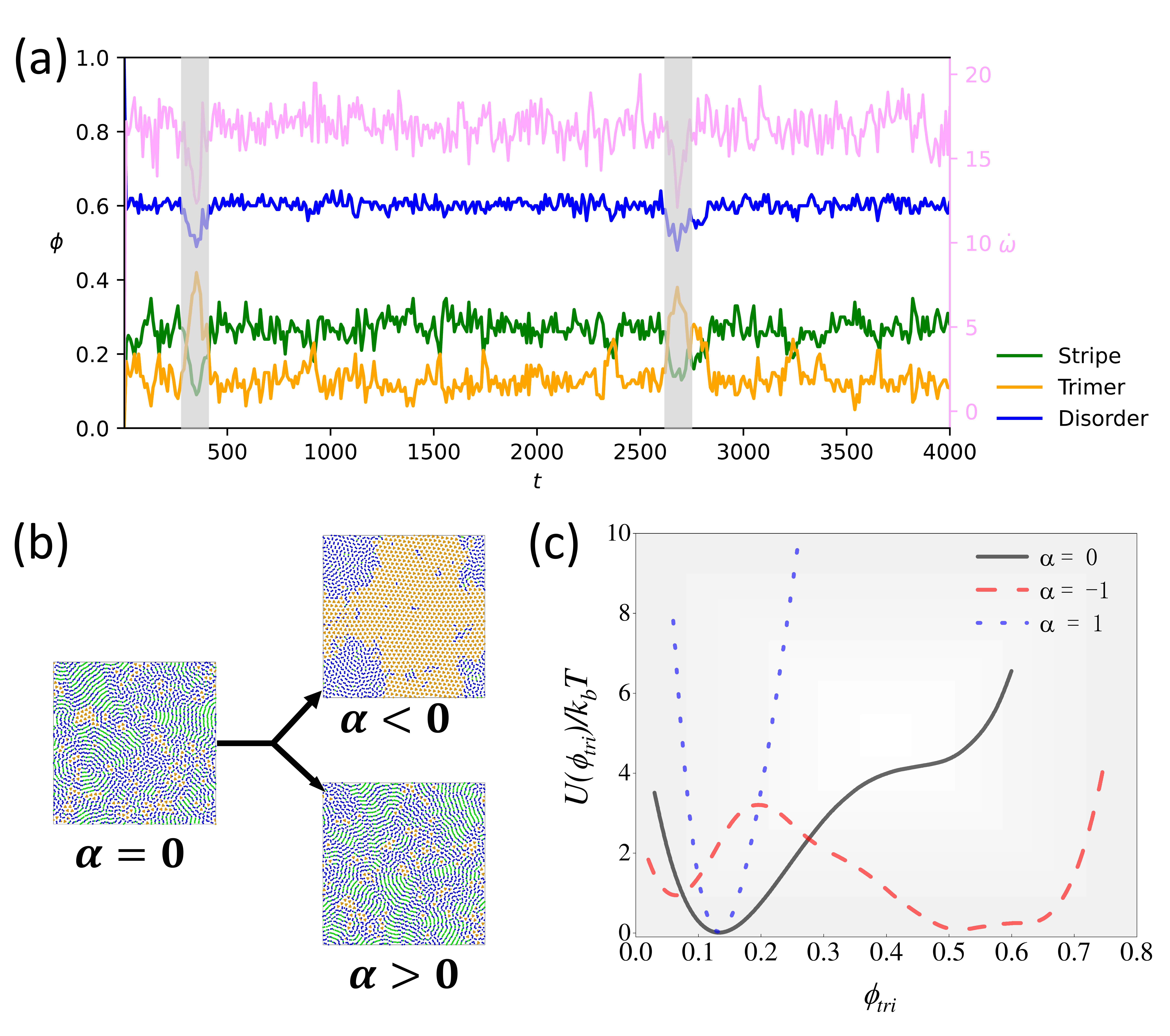}
\caption{(a) The temporal emergence of an ordered trimer cluster from a disordered structure is accompanied by a sudden drop in dissipation, highlighted by the grey-shaded region. We show the time series of the yields of the three states together with the active power input. (b) Representative simulation snapshots for different $\alpha$. Middle-left: no control of the dissipation tendency ($\alpha=0$). Upper-right: energy-avoiding pathways ($\alpha=-1$). Lower-right: energy-seeking pathways ($\alpha=1$). Parameters: $v_0=70$ and $k_s=190$. (c) The effective energy profiles along the reaction coordinate $\phi_{\mathrm{tri}}$. Black solid line: no control. Red dashed line: $\alpha=-1$. Blue dotted line: $\alpha=1$.}
\label{fig:3} 
\end{figure}

\textit{Induce ordered target configurations from the disordered structure.}--- In the following, we will show how the dissipation bias principle guides complex self-assembly processes in two representative scenarios. In the first scenario, we show that controlling the dissipation tendency can generate ordered trimer structures from disordered phases while improving their stability. We focus on the parameters with $v_0=70$ and $k_s=190$, a regime in which the uncontrolled dynamics ($\alpha=0$) remain largely disordered. At $\alpha=0$, the trimer yield remains low $\sim 0.1$, although trimer-rich local rearrangement clusters intermittently emerge, with yields reaching up to $\sim 0.4$. Interestingly, the dissipation drops precisely when trimer order emerges, indicating that the dissipation tendency provides a direct control handle (see Fig.~\ref{fig:3}a). Following our proposed dissipation bias principle, we set a negative  $\alpha$ ($\alpha=-1$) to suppress the system’s dissipation tendency to stabilize the emergent trimer structures. By repeatedly iterating the procedure shown in Fig.~\ref{fig:2}, the ensemble becomes progressively enriched through energy-avoiding pathways, which in turn steer the system toward an ordered configuration. In this regime, suppressing dissipation tendencies gives rise to additional energy punishment, which leads to an emergent spinodal stability. Therefore, trimer-rich clusters with lower dissipation can persist, leading to a marked increase in trimer yield, which can reach $\sim 0.7$ for $\alpha<0$ (see Fig.~\ref{fig:3}b and Fig.~S4). In contrast, a positive bias ($\alpha>0$) increases the dissipation tendency and promotes more dissipative dynamics, thereby suppressing the assembly of the trimer configurations (see Fig.~S4).
%Importantly, suppressing the dissipation tendency also reduces excessive fluctuations around the target state, consistent with a lower frequency and intensity of self-propulsion--driven local rearrangements. 

To further clarify our findings, we construct an effective energy landscape  \(U(\phi_{\mathrm{tri}})=-k_{\mathrm{B}}T\ln P_{\mathrm{ss}}(\phi_{\mathrm{tri}})\) from steady-state trajectory data by estimating \(P_{\mathrm{ss}}(\phi_{\mathrm{tri}})\) (see SM for details).  As shown in Fig.~\ref{fig:3}c, this representation makes the impact of dissipation-tendency control explicit. At \(\alpha=0\), \(U(\phi_{\mathrm{tri}})\) displays a single stable minimum near \(\phi_{\mathrm{tri}}\simeq 0.1\)--\(0.2\). When the dissipation tendency is suppressed (\(\alpha=-1<0\)), an additional basin can be observed at the high-$\phi_{\mathrm{tri}}$ side ($\phi_{\mathrm{tri}}\simeq 0.6$), indicating the stabilization of trimer-rich structures. In contrast, enhancing the dissipation tendency (\(\alpha=1>0\)) deepens and steepens the low-\(\phi_{\mathrm{tri}}\) basin, thereby suppressing the emergence of trimer-rich clusters.

\begin{figure}[htbp]
\includegraphics[width=0.98\columnwidth]{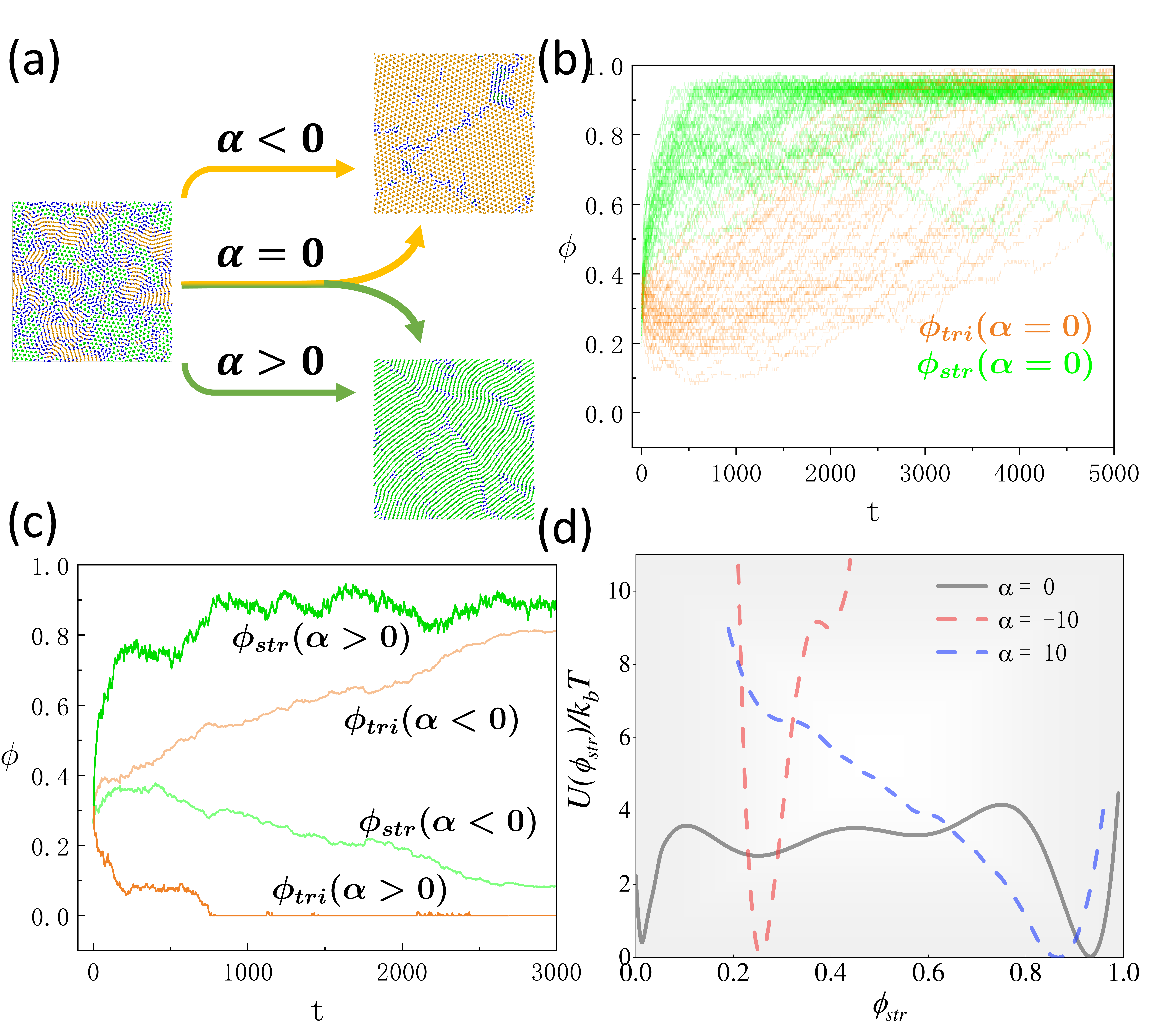}
\caption{(a) Representative snapshots from simulations for different $\alpha$. When there is no control of the dissipation tendency with $\alpha=0$, the system will assemble into both stripe-ordered configurations and trimer-ordered configurations through two different assembly pathways. On the one hand, for $\alpha=-10<0$, the energy-avoiding pathways will be collected. Particles directionally assemble into the trimer state with low dissipation. On the other hand, for $\alpha=10>0$, the energy-seeking path will be collected. Particles assemble into the stripe structure with high dissipation. (b) The time series of $\phi_{\mathrm{str}}$ (green) and $\phi_{\mathrm{tri}}$ (orange) measured along 100 independent trajectories with $\alpha=0$ . Each curve corresponds to one trajectory.  The parameters: $v_0=40$ and $k_s=220$.(c) The time series of the yields of the trimer structure (orange) and the stripe structure (green). Values of $\alpha$: $\alpha=10$ (light) and $\alpha=-10$ (dark). The parameters: $v_0=40$ and $k_s=220$. (d)The effective energy profiles of the system over the reaction coordinate $\phi_{str}$. Black solid line: no control. Red dashed line: $\alpha=-10$. Blue dotted line: $\alpha=10$.}  
\label{fig:4} 
\end{figure}

\textit{Directionally Selecting Among Multiple Assembly Pathways.}--- In the second scenario, we demonstrate that controlling dissipation tendencies provides an effective route to directed assembly. Multiple assembly pathways can coexist under the same conditions, resulting in distinct long-lived ordered outcomes, as observed in nature, experiments, and simulations \cite{ozbudak2004multistability,li2016dynamics}.  How to achieve directed assembly in multiple pathways is a challenge. Systems with competing dynamical pathways are sensitive to fluctuations, meaning that even identical initial conditions can lead to different ordered structures, a phenomenon indicative of replica symmetry breaking \cite{mezard1987spin}. At $v_0=40$ and $k_s=220$, simulations of 100 trajectories starting from random initial configurations resulted in either trimer or stripe configurations, which occurred with nearly equal likelihood (51 stripes versus 49 trimers, see Fig.~\ref{fig:4}b). Both configurations serve as stable attractors for the system, and transitions between stable states are not observed within the simulation's timescale. Interestingly, We quantify dissipation by calculating the active power input and find a distinct hierarchy: the dissipation associated with the stripe configuration is greater than that of the trimer configuration (see Fig.~S3). The significant difference in dissipative behavior between the stripe and trimer conformations provides us with a direct control handle. Following our proposed dissipation bias principle, we achieved directional self-assembly of the stripe and trimer configurations (see Fig.~\ref{fig:4}a). We set $\alpha<0$ to suppress the system’s dissipation tendency, thereby steering it along an energy-avoiding pathway that inhibits the formation of the highly dissipative stripe conformation and ultimately promotes the low-dissipation trimer configuration. In contrast, when \(\alpha>0\) is applied to enhance dissipation, the system follows an energy-seeking pathway and eventually converges to the stripe configuration. Once dissipation tendency control is introduced, the originally bistable assembly routes collapse into a single, directional pathway that can reliably connect to the targeted configuration. The time series of the yields of the trimer and stripe structures are shown in Fig.~\ref{fig:4}c.

%The self-assembly may have multiple pathways, linking to different stable states under given parameters, which have been observed in nature, experiments, and simulations \cite{ozbudak2004multistability,li2016dynamics}. How to achieve directed assembly in multiple pathways is a challenge. In the second scenario, we will demonstrate how dissipation control addresses this issue. Systems with multiple pathways are highly sensitive to fluctuations in the environment, leading the system to fall into different attractors. Taking $v_0=40$ and $k_s=220$ as an example, the system starting from random configurations will form different structures. We simulate over $100$ trajectories from random configurations, see \textcolor{red}{Fig.\ref{fig:4}b}. We observe that the system can assemble into both stripe-ordered configurations and trimer-ordered configurations through two different assembly pathways, as shown in \textcolor{red}{Fig.\ref{fig:4}a}. 

To further clarify our findings, we reconstruct the effective energy landscape $U(\phi_{\mathrm{str}})$. At $\alpha=0$, $U(\phi_{\mathrm{str}})$ exhibits a bistable profile with two well-separated basins: a low-$\phi_{\mathrm{str}}$ basin around $\phi_{\mathrm{str}}\!\approx\!0.05$ and a high-$\phi_{\mathrm{str}}$ basin around $\phi_{\mathrm{str}}\!\approx\!0.9$, consistent with two long-lived attractors and hence competing assembly pathways. Suppressing the dissipation tendency ($\alpha=-10<0$) strongly favors the low-$\phi_{\mathrm{str}}$ basin while destabilizing the high-$\phi_{\mathrm{str}}$ basin, whereas enhancing it ($\alpha=10>0$) instead stabilizes the high-$\phi_{\mathrm{str}}$ basin and penalizes low $\phi_{\mathrm{str}}$. In both cases, the control effectively tilts the landscape to remove pathway competition, yielding a predominantly single-funnel route toward the target configuration.

\textit{Summary and Discussion}—--We construct a dissipation-biased ensemble of nonequilibrium assembly trajectories within a large-deviation framework and realize it numerically using a cloning algorithm that periodically reweights trajectories by their active power. In this framework, the dissipation tendency becomes an explicit control parameter. It acts directly on the kinetic bottleneck of local rearrangements by modulating both their occurrence rate and typical strength, thereby reshaping the effective energy landscape in a way that cannot be achieved by static interaction design alone. With rapid advances in control technologies increasingly enabling external energy injection to be programmed in both space and time, regulating the dissipation tendency is becoming experimentally feasible. Although energy injection does not always increase a system's dissipation tendency, it nevertheless provides a practical handle for tuning the dissipation tendency. For example, the dissipation tendency can be tuned by modulating ATP concentration in biological systems \cite{england2015dissipative,te2018dissipative,perunov2016statistical,cao2024motorized,cao2025motorized} or by controlling pH in chemical systems to regulate chemical energy input \cite{tagliazucchi2014dissipative}; dissipation can also be regulated through designed reactions and recombination/mutation-like processes in building blocks \cite{zwicker2017growth,te2018dissipative}. Active colloids provide a particularly direct platform for controllable self-assembly in which local energy injection can be engineered to tune the dissipation tendency \cite{aubret2018targeted,mallory2019activity,mallory2018active,dey2016catalytic,aubret2017eppur,gao2017dynamic,singh2017non,lin2017light,theurkauff2012dynamic,pohl2014dynamic,jiang2010active,wang2013catalytically,ahmed2014self}, while nonequilibrium DNA systems offer a complementary route in which adaptive and programmable structures can be achieved by synchronizing energy-seeking/avoiding events with bond formation or cleavage \cite{heinen2019programmable}.

\begin{acknowledgments}
This work was supported by MOST (2022YFA1303100) and NSFC (22533005). 
\end{acknowledgments}

\bibliographystyle{apsrev4-2}
\bibliography{bibfile}

\end{document}